\pdfoutput=1
\documentclass[10pt, journal, rgb]{IEEEtran}

\usepackage[usenames,dvipsnames,table]{xcolor}
\usepackage{amsmath,amssymb,amsfonts,amsthm}
\usepackage[pdftex]{graphicx}

\usepackage[T1]{fontenc}
\usepackage[utf8]{inputenc}

\usepackage{listings}
\usepackage{csquotes}
\usepackage{nicefrac}
\usepackage[textsize=tiny]{todonotes}
\usepackage{booktabs}
\usepackage[hyphens,spaces,obeyspaces]{url}
\usepackage[bookmarks=false]{hyperref}
\usepackage{tikz}
\usetikzlibrary{shapes.geometric, positioning, calc, matrix, fit}

\usepackage{physics}
\usepackage{yquant}

\makeatletter
\let\MYcaption\@makecaption
\makeatother
\usepackage[font=footnotesize]{subcaption}
\makeatletter
\let\@makecaption\MYcaption
\makeatother

\usepackage[style=ieee, maxnames=4, minnames=1, date=year, doi=false,isbn=false,backend=biber]{biblatex}
\usepackage[binary-units=true, detect-all=true]{siunitx}
\usepackage{etoolbox}
\robustify\bfseries

\usepackage{nicematrix}
\usepackage{flushend}

\newtheorem{example}{Example}

\tikzset{%
	font={\footnotesize},
	terminal/.style={draw,rectangle,inner sep=2pt,font=\footnotesize,very thick},
	vertex/.style={draw,circle,inner sep=0pt,minimum width=0.5cm,minimum height=0.5cm},
	define color/.code={\definecolor{hsb#1}{Hsb}{#1, 1, 0.75}},				
    medge/.style n args={3}{
		line width={#1pt},
		define color={#2},
		draw=hsb#2,
		out=#3, 
		in=90
	},
    edge/.style 2 args={
		line width={#1pt},
		define color={#2},
		draw=hsb#2
	},
	edge0/.style 2 args={
		line width={#1pt},
		define color={#2},
		draw=hsb#2,
		out=-130, 
		in=90
	},
	edge1/.style 2 args={
		line width={#1pt},
		define color={#2},
		draw=hsb#2,
		out=-50, 
		in=90
	},
	zerostub/.style={
		inner sep=0, 
		minimum size=3pt, 
		circle, 
		fill=black
	}
}

\begin{document}

\title{\huge Simulation Paths for \\Quantum Circuit Simulation with Decision Diagrams \\{\vspace{-3mm}\Large What to Learn from Tensor Networks, and What Not}}

\author{
	\IEEEauthorblockN{Lukas Burgholzer\IEEEauthorrefmark{1}\hspace*{1.5cm}Alexander Ploier\IEEEauthorrefmark{1}\hspace*{1.5cm}Robert Wille\IEEEauthorrefmark{2}\IEEEauthorrefmark{3}} \\
	\IEEEauthorblockA{\IEEEauthorrefmark{1}Institute for Integrated Circuits, Johannes Kepler University Linz, Austria}\\
	\IEEEauthorblockA{\IEEEauthorrefmark{2}Chair for Design Automation, Technical University of Munich, Germany}\\
	\IEEEauthorblockA{\IEEEauthorrefmark{3}Software Competence Center Hagenberg GmbH (SCCH), Austria}\\
	\IEEEauthorblockA{\href{mailto:lukas.burgholzer@jku.at}{lukas.burgholzer@jku.at}\hspace{1.5cm}\href{mailto:alexander.ploier@jku.at}{alexander.ploier@jku.at}\hspace{1.5cm} \href{mailto:robert.wille@jku.at}{robert.wille@jku.at}\\
	\url{https://www.cda.cit.tum.de/research/quantum/}}
	\vspace*{-1cm}
}

\maketitle

\begin{abstract}
Simulating quantum circuits on classical computers is a notoriously hard, yet increasingly important task for the development and testing of quantum algorithms.
In order to alleviate this inherent complexity, efficient data structures and methods such as tensor networks and decision diagrams have been proposed.
However, their efficiency heavily depends on the order in which the individual computations are performed. For tensor networks the order is defined by so-called \emph{contraction plans} and a plethora of methods has been developed to determine suitable plans. 
On the other hand, simulation based on decision diagrams is mostly conducted in a straight-forward, i.e., sequential, fashion thus far.

In this work, we study the importance of the path that is chosen when simulating quantum circuits using decision diagrams and show, conceptually and experimentally, that choosing the right simulation path can make a vast difference in the efficiency of classical simulations using decision diagrams.
We propose an open-source framework (available at \url{github.com/cda-tum/ddsim}) that not only allows to investigate dedicated simulation paths, but also to re-use existing findings, e.g., obtained from determining contraction plans for tensor networks.
Experimental evaluations show that translating strategies from the domain of tensor networks may yield speedups of several factors compared to the state of the art. 
Furthermore, we design a dedicated simulation path heuristic that allows to improve the performance even further---frequently yielding speedups of several orders of magnitude.
Finally, we provide an extensive discussion on what \emph{can} be learned from tensor networks and what \emph{cannot}.
\end{abstract}

\section{Introduction}
\label{sec:introduction}

Over the last couple of years, quantum computers have evolved from a theoretical computational model to practical devices aimed for pushing beyond the horizon of classically tractable problems.
Even though actual quantum computers have already been built, their availability is still rather limited and devices are heavily affected by noise.
Moreover, in order to develop and test potential applications, complete representations of the respective quantum states are needed---information which is fundamentally unavailable on actual devices.
Hence, the simulation of quantum computations on classical machines plays a vital role in the ongoing race to realize useful applications for quantum computing.

Such simulations entail computing a representation of the state resulting from the application of a sequence of operations (typically described as a quantum circuit) to the initial state of a quantum system.
Conceptually, this corresponds to a sequence of matrix-vector multiplications.
While simple in principle, the underlying vectors and matrices grow exponentially with respect to the number of simulated qubits (the quantum analogue to the bit)---quickly requiring powerful supercomputing clusters to feasibly conduct the classical simulations~\cite{guerreschiIntelQuantumSimulator2020, hanerPetabyteSimulation45Qubit2017, jonesQuESTHighPerformance2018}.
Clever data structures such as tensor networks~\cite{jozsaSimulationQuantumCircuits2006, markovSimulatingQuantumComputation2008, biamonteTensorNetworksNutshell2017} or decision diagrams~\cite{niemannQMDDsEfficientQuantum2016, chin-yungExtendedXQDDRepresentation2011, zulehnerHowEfficientlyHandle2019,hongTensorNetworkBased2020} have been demonstrated to alleviate this complexity in many practically relevant cases. 

To translate the problem of classically simulating a quantum circuit into the tensor network domain, each gate of the circuit as well as the initial state is represented as a tensor and individual tensors are connected via shared indices.
Then, simulating the quantum circuit entails contracting all connected tensors along the shared indices until only a single tensor remains.
It is commonly known that the complexity of such a simulation is extremely sensitive to the order in which the individual tensors are contracted. 
Accordingly, a plethora of methods have been developed to efficiently determine suitable contraction paths~\cite{grayHyperoptimizedTensorNetwork2021, huangClassicalSimulationQuantum2020, boixoSimulationLowdepthQuantum2018, lykovTensorNetworkQuantum2020}---a task proven to be NP-hard~\cite{chi-chungOptimizingClassMultidimensional1997}.
The general idea of such contraction plans is to exploit the topological structure of the quantum circuit.

Decision diagrams, on the other hand, try to compactly represent the individual operations and quantum states by exploiting redundancies in the underlying representation. 
To this end, they represent these quantities as directed, acyclic graphs with complex-valued edge weights.
Similar to tensor networks, the initial quantum state and each gate of a quantum circuit are first translated to their (typically \mbox{linearly-sized}) decision diagram representation.
Simulation of such a system is then conducted by multiplying the respective decision diagrams until a decision diagram representation of the final state vector remains.
Therein, the complexity of multiplying decision diagrams scales with the product of their sizes, i.e., their number of nodes.
Whenever the respective intermediate decision diagrams remain rather compact, an efficient scheme for classical simulation is obtained.

Therefore, data structures such as tensor networks and decision diagrams can help to alleviate the complexity of simulating quantum circuits. 
In both cases, the efficiency heavily depends on the order in which computations are performed, namely the contraction plan for tensor networks and the order of matrix-matrix or matrix-vector multiplications for decision diagrams---called \mbox{\emph{simulation path}} in the following.
While thoroughly investigated for tensor networks, this effect has hardly been studied for decision diagrams yet.

In this work\footnote{A preliminary version of this work has been published in~\cite{burgholzerExploitingArbitraryPaths2022}.}, we investigate this issue
and propose an \mbox{open-source} framework that allows to exploit arbitrary simulation paths for decision diagram-based quantum circuit simulation.
Instead of reinventing the wheel, we establish a flow that not only allows to investigate dedicated paths but also to re-use existing techniques, e.g., from the tensor network domain, also for decision diagrams.

Considering the verification of quantum circuit compilation flow results as a particularly important use case for quantum circuit simulation, we show, both conceptually and experimentally, that choosing the right simulation path can make a vast difference in the efficiency of classical simulations using decision diagrams.
To this end, we demonstrate that translating strategies from the domain of tensor networks allows for speedups of several factors compared to the state of the art in many cases.
Additionally, we design a dedicated simulation path heuristic that allows to improve the performance even further---frequently yielding speedups of several orders of magnitude.
Based on these evaluations, we discuss the resulting consequences on what \emph{can} be learned from tensor networks and what \emph{cannot} be learned from them.
This eventually provides the basis for future research on quantum circuit simulation using decision diagrams.

The rest of this paper is structured as follows: \autoref{sec:background} introduces the necessary background for the rest of this work.
Then, \autoref{sec:motivation} illustrates the degrees of freedom and the potential impact of arbitrary simulation paths. Motivated by that, \autoref{sec:framework} presents the framework that allows to evaluate these arbitrary simulation paths and describes how existing techniques from the tensor network domain can be used to obtain \enquote{good} simulation paths without starting from scratch. 
Afterwards, \autoref{sec:results} summarizes our experimental evaluations, followed by a discussion of their implications in \autoref{sec:discussion}.
\autoref{sec:conclusions} concludes the paper.

\section{Background}
\label{sec:background}

To keep this paper self-contained, this section briefly covers the basics on quantum circuit simulation followed by a brief review of decision diagrams---which provide the basis of the simulation approach considered in the rest of this work.

\subsection{Quantum Circuit Simulation}
\label{sec:quantum_computing}

A quantum state $\ket{\varphi}$ of an $n$-qubit quantum system can be described as a linear combination of $2^n$ basis states, i.e., 
\begin{equation*}
\ket{\varphi} = \sum_{i\in\{0,1\}^n} \alpha_i \ket{i} \mbox{ with } \alpha_i\in\mathbb{C} \mbox{ and }\sum_{i\in\{0,1\}^n} \vert\alpha_i\vert^2 = 1.
\end{equation*}
This state is commonly represented as a vector~$[\alpha_{0\dots 0}, \dots, \alpha_{1\dots 1}]^\top$, referred to as \emph{state vector}. Measuring this state leads to a collapse of the system's state to one of the basis states~$\ket{i}$---each with probability $\vert \alpha_i\vert^2$ for $i\in\{0,1\}^n$. In the following, we will always identify $\ket{\varphi}$ with its corresponding state vector, i.e., 
\begin{equation*}
\ket{\varphi}\equiv[\alpha_{0\dots 0}, \dots, \alpha_{1\dots 1}]^\top.
\end{equation*}

\begin{example}\label{ex:state}
An important example of a quantum state is the $3$-qubit Greenberger–Horne–Zeilinger or GHZ state~\cite{nielsenQuantumComputationQuantum2010}: 
\[
\ket{\mathit{GHZ}} = \frac{1}{\sqrt{2}} (\ket{000} + \ket{111}) \equiv \frac{1}{\sqrt{2}}[1,0,0,0,0,0,0,1]^\top
\]
\end{example}

The state of any quantum system can be manipulated by quantum operations, also called \emph{quantum gates}. Any such gate~$g$ acting on $k$ qubits can be identified by a unitary matrix~$U$ of size $2^k \times 2^k$, i.e., $g \equiv U$. 
Its action on a quantum state corresponds to the matrix-vector product of the matrix with the vector representing the state\footnote{Technically, the matrix first needs to be extended to the full system size (by forming appropriate tensor products with identity matrices) for the multiplication to make sense.}, i.e., $\ket{\varphi'} = U \ket{\varphi}$.

A \emph{quantum circuit} is now described as a composition of quantum gates.
Consequently, the evolution of an initial quantum state $\ket{\varphi}$ through a quantum circuit $G = g_1 \dots g_{|G|}$ is described by the subsequent application of the individual gates to this initial state, i.e.,  
\[
\ket{\varphi} G  = \ket{\varphi} g_1 \dots g_{|G|} \equiv U_{|G|} * \cdots * U_1 * \ket{\varphi}.
\]
If this task is conducted on a classical computer, it is commonly referred to as \emph{quantum circuit simulation}.

\begin{figure}[t]
	\centering
	\resizebox{0.62\linewidth}{!}{
	\begin{tikzpicture}
	\begin{yquant}
		qubit {} q[3];
		h q[0];
		box {$S$} q[0] | q[1];
		box {$T$} q[0] | q[2];
		h q[1];
		box {$S$} q[1] | q[2];
		h q[2];
		swap (q[0], q[2]);
	\end{yquant}
\end{tikzpicture}
	}
	\caption{Circuit for the $3$-qubit quantum Fourier transform}
	\label{fig:circuit}
\end{figure}
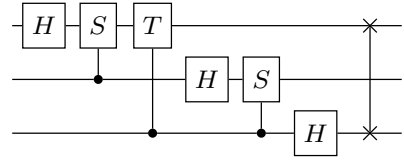
\begin{example}\label{ex:operation}
\autoref{fig:circuit} shows a $3$-qubit quantum circuit realizing an important quantum algorithm, namely the quantum analog to the Fourier transform.
It consists of three \mbox{single-qubit} Hadamard gates (indicated by boxes labeled $H$), three \mbox{two-qubit} controlled-phase rotations (indicated by boxes labeled $S$ and $T$ connected to $\bullet$), and a two-qubit SWAP gate (indicated by $\times$).
Given an initial state in the computational basis, this circuit outputs the state's representation in the Fourier basis.
\end{example}

\subsection{Decision Diagrams}
\label{sec:decision_diagrams}

The representations of a quantum system state and the operations manipulating it are exponentially large with respect to the number of qubits involved.
This quickly limits \mbox{straight-forward} approaches for representing (and manipulating) even moderately sized state vectors, such as arrays, without resorting to supercomputing clusters. For example, representing the dense state vector of a $32$-qubit system already requires \SI{64}{\gibi\byte} of memory (assuming \SI{128}{\bit} complex numbers).

\emph{Decision diagrams}~\cite{zulehnerHowEfficientlyHandle2019,chin-yungExtendedXQDDRepresentation2011,niemannQMDDsEfficientQuantum2016,hongTensorNetworkBased2020} have been proposed as a complementary approach for efficiently representing and manipulating quantum states by exploiting redundancies in the underlying representation.
A decision diagram representing a quantum state (or operation) is a directed, acyclic graph with complex edge weights.
To this end, a given state vector with its complex amplitudes \(\alpha_i\) for \(i \in \{0,1\}^n\) is recursively decomposed into sub-vectors according to
\begin{gather*}
        [\alpha_{0\ldots 0}, \hdots, \alpha_{1\ldots 1}]^\top \\
        [\alpha_{0x}]^\top \qquad\qquad\qquad [\alpha_{1x}]^\top \\
        [\alpha_{00y}]^\top \quad [\alpha_{01y}]^\top \qquad [\alpha_{10y}]^\top \quad [\alpha_{11y}]^\top,
\end{gather*}
with $x\in\{0,1\}^{n-1}$ and $y\in\{0,1\}^{n-2}$, until only individual amplitudes remain. The resulting graph has $n$ levels of nodes, labelled $n-1$ down to $0$. Here, each node $i$ has exactly two successors indicating whether the path leads to an amplitude where qubit $i$ is in state $\ket{0}$ or $\ket{1}$.

By extracting common factors into edge weights (and employing suitable normalization schemes, see~\cite{niemannQMDDsEfficientQuantum2016,zulehnerHowEfficientlyHandle2019}), any two sub-vectors that only differ by a constant factor can be unified and need not be represented by separate nodes in the decision diagram.
Exploiting such redundancies frequently allows to obtain rather compact representations 
(in the best case linear with respect to the number of qubits)  for the, in general, exponentially large state vectors. 

\begin{figure}
\centering
\begin{subfigure}[b]{.24\linewidth}
  \centering
  \resizebox{0.9\textwidth}{!}{
  \begin{tikzpicture}[node distance=1 and 0.5, 	on grid]
			\node[vertex] (q2) {2};
			\node[vertex, below left=of q2] (q1l) {1};
			\node[vertex, below right=of q2] (q1r) {1};
			\node[vertex, below=of q1l] (q0l) {0};
			\node[vertex, below=of q1r] (q0r) {0};
			\node[terminal, below right=of q0l] (t) {};
			
			\draw[edge={2}{0}] ($(q2)+(0,0.7cm)$) -- (q2);
			
			\draw[edge0={1}{0}] (q2) to (q1l);
			\draw[edge1={1}{0}] (q2) to (q1r);

			\draw[edge0={2}{0}] (q1l) to (q0l);
			\draw[black, thick] (q1l) -- ++(-50:0.35) node[zerostub]{};

			\draw[edge1={2}{0}] (q1r) to (q0r);
			\draw[black, thick] (q1r) -- ++(-130:0.35) node[zerostub]{};

			\draw[edge0={2}{0}] (q0l) to (t);
			\draw[black, thick] (q0l) -- ++(-50:0.35) node[zerostub]{};

			\draw[edge1={2}{0}] (q0r) to (t);
			\draw[black, thick] (q0r) -- ++(-130:0.35) node[zerostub]{};
	\end{tikzpicture}}
  \caption{$\ket{\mathit{GHZ}}$}
  \label{fig:basis_state_dd}
\end{subfigure}%
\begin{subfigure}[b]{0.26\linewidth}
\centering
\resizebox{0.99\linewidth}{!}{
\begin{tikzpicture}
	\begin{scope}[font=\large,
					   define color/.code={
					       \definecolor{hsb#1}{Hsb}{#1, 1, 0.75}
					   },
					   wheel color/.style={
					   		line width=4mm,
					   		define color={#1},
					   		draw=hsb#1
					   }]
					\foreach \x in {0,2,...,358} {
						\draw [wheel color=\x] (\x:0.9) arc (\x:\the\numexpr\x+3:0.9);
					}
					\node at (0:1.4) {\(0\)};
					\node at (90:1.4) {\(\frac{\pi}{2}\)};
					\node at (180:1.4) {\(\pi\)};
					\node at (270:1.5) {\(\frac{3\pi}{2}\)};
			\end{scope}
\end{tikzpicture}
}
\caption{Color wheel}
\label{fig:colorwheel}	
\end{subfigure}%
\begin{subfigure}[b]{0.28\linewidth}
\centering	
\resizebox{0.85\linewidth}{!}{
\begin{tikzpicture}[node distance=1 and 0.5, on grid]
		\node[vertex] (q2) {2};
		\node[vertex, below=of q2] (q1) {1};
		\node[vertex, below left=of q1] (q0l) {0};
		\node[vertex, below right=of q1] (q0r) {0};
		\node[terminal, below right=of q0l] (t) {};
		
		\draw[edge={2}{0}] ($(q2)+(0,0.7cm)$) -- (q2);
		
		\draw[medge={2}{0}{-130}] (q2) to (q1);
		\draw[black, thick] (q2) to ++(-100:0.35) node[zerostub]{};
		\draw[black, thick] (q2) to ++(-80:0.35) node[zerostub]{};
		\draw[medge={2}{0}{-50}] (q2) to (q1);
		
		\draw[medge={2}{0}{-130}] (q1) to (q0l);
		\draw[black, thick] (q1) to ++(-100:0.35) node[zerostub]{};
		\draw[black, thick] (q1) to ++(-80:0.35) node[zerostub]{};
		\draw[medge={2}{0}{-50}] (q1) to (q0r);
		
		\draw[medge={2}{0}{-130}] (q0l) to (t);
		\draw[black, thick] (q0l) to ++(-100:0.35) node[zerostub]{};
		\draw[black, thick] (q0l) to ++(-80:0.35) node[zerostub]{};
		\draw[medge={2}{0}{-50}] (q0l) to (t);
		
		\draw[medge={2}{0}{-130}] (q0r) to (t);
		\draw[black, thick] (q0r) to ++(-100:0.35) node[zerostub]{};
		\draw[black, thick] (q0r) to ++(-80:0.35) node[zerostub]{};
		\draw[medge={2}{90}{-50}] (q0r) to (t);
		\end{tikzpicture}
}
\caption{Controlled-S gate}
\label{fig:czdd}
\end{subfigure}%
\begin{subfigure}[b]{.18\linewidth}
	\centering
	\resizebox{0.5\linewidth}{!}{
 	\begin{tikzpicture}[on grid]
		\matrix[matrix of nodes,ampersand replacement=\&,column sep={1cm,between origins},row sep={1cm,between origins}] (qmdd) {
			\node[draw = none] (top) {}; \\
			\node[vertex] (n3) {2}; \\
			\node[vertex] (n2) {1}; \\
			\node[vertex] (n1) {0}; \\
			\node[terminal] (t){};\\
		};
		\draw[medge={2}{0}{-130}] (n3) to (n2);
		\draw[medge={2}{0}{-50}] (n3) to (n2);
		\draw[black, thick] (n3) to ++(-100:0.35) node[zerostub]{};
		\draw[black, thick] (n3) to ++(-80:0.35) node[zerostub]{};
		
		\draw[medge={2}{0}{-130}] (n2) to (n1);
		\draw[medge={2}{0}{-50}] (n2) to (n1);
		\draw[black, thick] (n2) to ++(-100:0.35) node[zerostub]{};
		\draw[black, thick] (n2) to ++(-80:0.35) node[zerostub]{};

		\draw[medge={2}{0}{-130}] (n1) to (t);
		\draw[medge={2}{0}{-100}] (n1) to (t);
		\draw[medge={2}{0}{-80}] (n1) to (t);
		\draw[medge={2}{180}{-50}] (n1) to (t);
		
		\draw[medge={1}{0}{-90}] (top) -- (n3);
		\end{tikzpicture}}  
		\caption{H gate}
  \label{fig:hadamard_dd}
\end{subfigure}%
\vspace*{-3mm}
\caption{Decision diagrams for $3$-qubit states and gates}
\label{fig:dds}
\vspace{-2mm}
\end{figure}
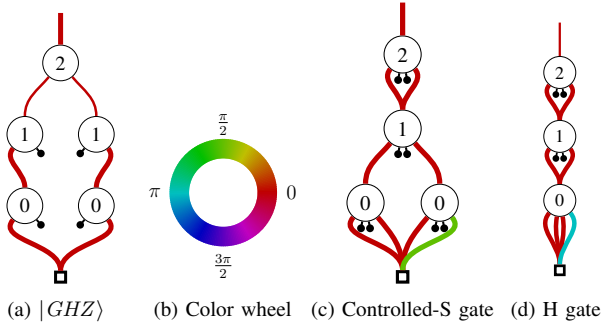

\begin{example}\label{ex:dd_state}
	\autoref{fig:basis_state_dd} shows a graphical representation (as proposed in~\cite{willeVisualizingDecisionDiagrams2021}) of the decision diagram for the GHZ state considered previously in \autoref{ex:state}.
	To this end, the thickness of an edge indicates the magnitude of the corresponding weight, while the color wheel shown in \autoref{fig:colorwheel} is used to encode its phase.
	Furthermore, edges with a weight of $0$ are denoted as $\bullet$-stubs. 
	In general, the decision diagram for an $n$-qubit GHZ state requires $2n-1$ nodes for representing the \mbox{$2^n$-dimensional} state vector.
\end{example}

Decision diagram representations for quantum gates are obtained by extending the decomposition scheme for state vectors by a second dimension.
This corresponds to recursively splitting the respective matrix into four equally sized \mbox{sub-matrices} according to the basis
\[
\left\{\begin{bNiceMatrix}1&0\\0&0\end{bNiceMatrix}, \begin{bNiceMatrix}0&1\\0&0\end{bNiceMatrix}, \begin{bNiceMatrix}0&0\\1&0\end{bNiceMatrix}, \begin{bNiceMatrix}0&0\\0&1\end{bNiceMatrix}\right\}.
\]

\begin{example}\label{ex:dd_example}
Consider again the circuit shown in \autoref{ex:operation}. 
Then, \autoref{fig:czdd} and \autoref{fig:hadamard_dd} show the decision diagram representations for the $2^3\times 2^3$ matrices of the controlled-S  and the Hadamard gate at the end of the circuit, respectively.
\end{example}

As described above, applying a gate to a quantum system entails the matrix-vector multiplication of the corresponding matrix with the current state vector.
This operation can be recursively broken down according to
\begin{align*}
\begin{bmatrix}
U_{00} & U_{01} \\
U_{10} & U_{11} \\
\end{bmatrix}
*
\begin{bmatrix}
\alpha_{0\ldots} \\
\alpha_{1\ldots} \\
\end{bmatrix}
=
\begin{bmatrix}
(U_{00} * \alpha_{0\ldots} + U_{01} * \alpha_{1\ldots}) \\
(U_{10} * \alpha_{0\ldots} + U_{11} * \alpha_{1\ldots}) \\
\end{bmatrix},
\end{align*}
with $U_{ij}\in\mathbb{C}^{2^{n-1}\times 2^{n-1}}$ and $\alpha_{i\ldots}\in\mathbb{C}^{2^{n-1}}$ for $i,j\in\{0,1\}$.
Since the $U_{ij}$ and $\alpha_{i\ldots}$ directly correspond to the successors in the respective decision diagrams, matrix-vector (as well as matrix-matrix) multiplication is a native operation on decision diagrams and its complexity scales with the product of the number of nodes of both decision diagrams.
Thus, whenever the decision diagrams remain compact throughout the computation, the simulation of quantum circuits can be efficiently conducted using decision diagrams~\mbox{\cite{zulehnerAdvancedSimulationQuantum2019, samoladasImprovedBDDAlgorithms2008, viamontesHighperformanceQuIDDBasedSimulation2004, hillmichAccurateNeededEfficient2020}}.
While many practical examples lead to compact decision diagram representations~\cite{grurlArraysVsDecision2020}, their worst case complexity remains exponential.

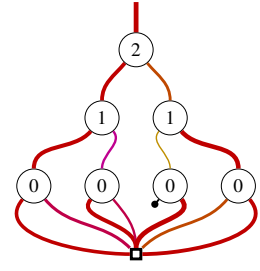
\begin{figure}[t]
	\centering
  \resizebox{0.45\linewidth}{!}{
  \begin{tikzpicture}[node distance=1 and 0.5, 	on grid]
			\node[vertex] (q2) {2};
			\node[vertex, below left=of q2] (q1l) {1};
			\node[vertex, below right=of q2] (q1r) {1};
			\node[vertex, below=of q1l] (q0l) {0};
			\node[vertex, below=of q1r] (q0r) {0};
			\node[vertex, left= 1 of q0l] (q0ll) {0};
			\node[vertex, right= 1 of q0r] (q0rr) {0};
			\node[terminal, below right=of q0l] (t) {};
			
			\draw[edge={2}{0}] ($(q2)+(0,0.7cm)$) -- (q2);
			
			\draw[edge0={1.6}{0}] (q2) to (q1l);
			\draw[edge1={1.2}{22.5}] (q2) to (q1r);

			\draw[edge0={1.7}{0}] (q1l) to (q0ll);
			\draw[edge1={1}{315}] (q1l) to (q0l);

			\draw[edge1={1.9}{0}] (q1r) to (q0rr);
			\draw[edge0={0.6}{45}] (q1r) to (q0r);

			\draw[edge0={1.7}{0}] (q0l) to (t);
			\draw[edge1={1}{337.5}] (q0l) to (t);

			\draw[black, thick] (q0r) -- ++(50:-0.35) node[zerostub]{};
			\draw[edge1={2}{0}] (q0r) to (t);
			
			\draw[edge0={1.2}{22.5}] (q0rr) to[in=45] (t);
			\draw[edge1={1.6}{0}] (q0rr) to[in=0] (t);
			
			\draw[edge0={1.5}{0}] (q0ll) to[in=180] (t);
			\draw[edge1={1.3}{337.5}] (q0ll) to[in=135] (t);
	\end{tikzpicture}}
  \caption{Maximally large decision diagram resulting from $\mathit{QFT}\ket{\mathit{GHZ}}$}
  \label{fig:wc_dd}
  \vspace*{-3mm}
\end{figure}

\begin{example}\label{ex:worstcasedd}
	Let $\ket{\varphi}$ denote the $n$-qubit GHZ state and let $G$ be the circuit for the $n$-qubit quantum Fourier transform.
	As demonstrated in \autoref{ex:dd_state} and \autoref{ex:dd_example} for $n=3$, both, the decision diagram representations for the initial state $\ket{\varphi}$ as well as the individual gates are linear.
	However, it can be shown that the decision diagram of the final state resulting from the simulation of $G$ with initial state $\ket{\varphi}$ is maximally large, i.e., consists of $2^n-1$ nodes. \autoref{fig:wc_dd} shows the corresponding decision diagram for $n=3$.
\end{example}

\section{Motivation and Related Work}
\label{sec:motivation}
In this section, we consider the question of how the order in which the respective multiplications are conducted influences the complexity of decision diagram-based simulation---a topic hardly considered thus far.
Afterwards, we discuss correspondingly related work including how other types of quantum circuit simulators address this problem.

\begin{figure*}
\centering
\includegraphics[width=0.7\linewidth]{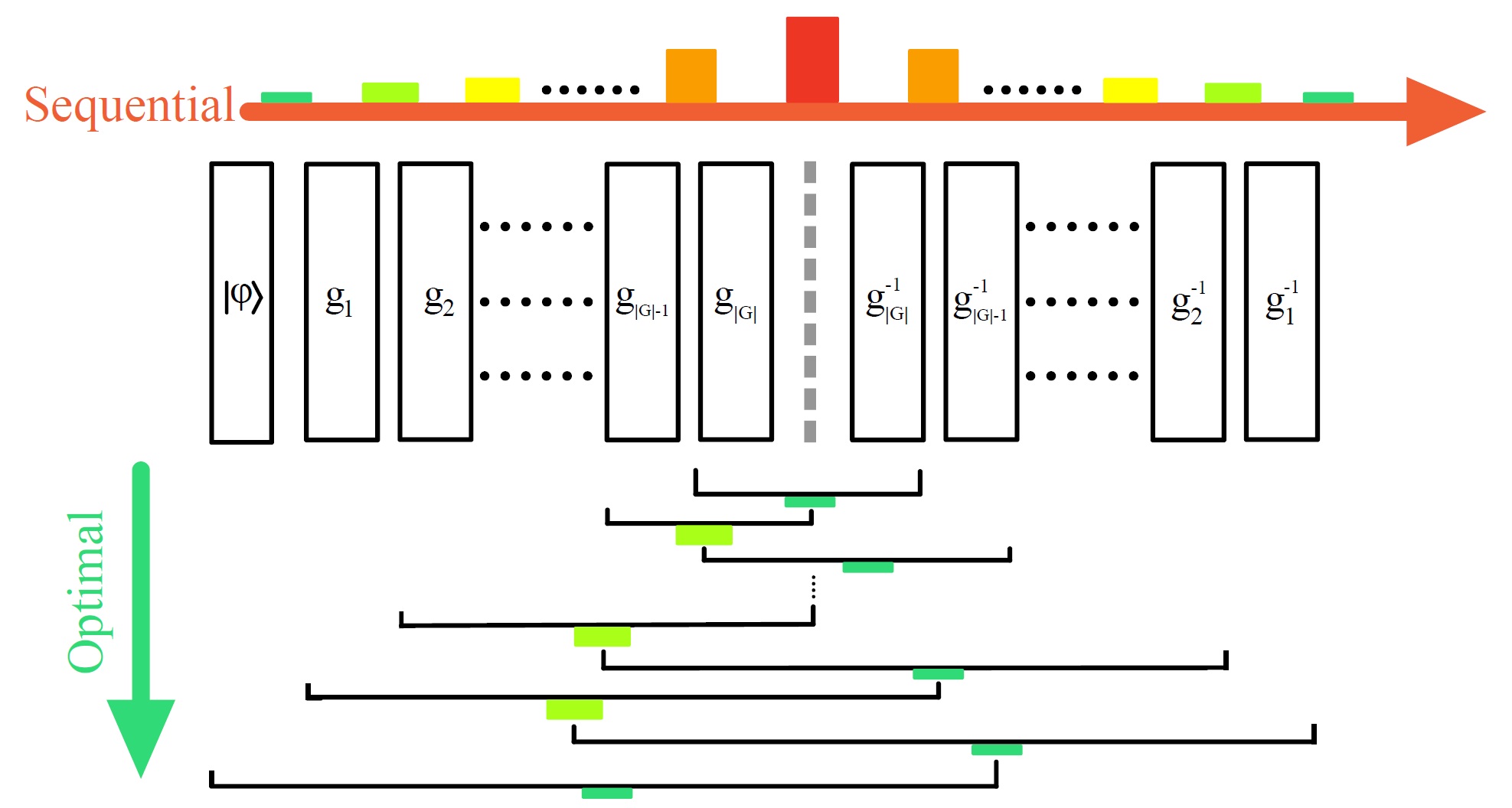}	
\caption{Comparison of the sequential and the optimal simulation path for the problem of verifying the equivalence of two quantum circuits}\label{fig:optipath}
\end{figure*}

\subsection{Considered Problem}
\label{sec:considered_problem}

As reviewed above, the simulation of a quantum circuit \mbox{$G = g_1 \dots g_{|G|}$} given an initial state $\ket{\varphi}$ entails the sequence of computations
\[
\ket{\varphi} G  = \ket{\varphi} g_1 \dots g_{|G|} \equiv U_{|G|} * \cdots * U_1 * \ket{\varphi}.
\]
Since matrix-matrix and matrix-vector multiplication is associative, the order in which the individual multiplications are conducted can, in principle, be chosen arbitrarily.
We refer to such an order of computations as a \emph{simulation path}.
Due to matrix-vector multiplication, in general, being far less complex than matrix-matrix multiplication, the most natural simulation path is to \emph{sequentially} compute the matrix-vector product of the individual (and compact) gate matrices with the current state vector. 
However, for a circuit $G$ with $|G|$ gates, there are
\begin{equation*}
|G|*(|G|-1)*\cdots * 1= |G|!,
\end{equation*} i.e., exponentially many, unique simulation paths---raising the question whether the most natural path indeed is always the best path.

In order to demonstrate the impact of the simulation path on the simulation complexity, we consider the following typical  use case for quantum circuit simulation:
Given two quantum circuits $G = g_1 \dots g_{|G|}$ and $G' = g'_1\dots g'_{|G'|}$, it shall be checked whether both circuits are equivalent---an essential question when, e.g., verifying the results of quantum circuit compilation flows~\cite{burgholzerVerifyingResultsIBM2020}.
Due to quantum circuits being inherently reversible, this can be checked by concatenating one circuit with the inverse of the other, i.e.,
\[
\tilde{G} = G G^{\prime -1} = g_1 \dots g_{|G|}\; g_{|G'|}^{\prime -1} \cdots g_{1}^{\prime -1},
\]
and simulating the resulting circuit with various initial \mbox{states $\ket{\varphi}$}.
Whenever $G$ and $G'$ are equivalent, $\tilde{G}\equiv\mathit{I}$ holds (with $I$ denoting the identity transformation) and, hence, $\tilde{G}$ maps $\ket{\varphi}$ to itself.
However, as the following example will show, choosing the right simulation path for the simulation of $\tilde{G}$ can make the difference between linear and exponential complexity.

\begin{example}\label{ex:worstcasesimpath}
Consider the scenario as in \autoref{ex:worstcasedd} and, for the sake of the argument, assume that \mbox{$G' = G$}, i.e., it naturally holds $\tilde{G}=G G^{-1} \equiv I$ for any $\ket{\varphi}$.
Then, following the discussion in \autoref{ex:worstcasedd}, simulating $\tilde{G}$ in a sequential fashion leads to an intermediate decision diagram that is maximally large---implying an exponential memory complexity and, hence, exponential runtime.
If however, the simulation path is chosen to start \enquote{in between} $G$ and $G^{-1}$ and \emph{alternate} between applying gates from $G$ and $G^{-1}$, any computation (except the last \mbox{matrix-vector} multiplication) has the form
\[
U^{-1}_i * U_i \quad\mbox{ or }\quad I * U_i,
\]
for $i=1,\dots,|G|$.
Since, in general, the complexity of decision diagrams representing individual gates is linear, the overall runtime and memory complexity is linear as well.
\autoref{fig:optipath} illustrates this scenario, i.e., sketches the respectively applied matrix-matrix or matrix-vector multiplications, in addition to a color palette indicating the size of the correspondingly resulting decision diagrams (with green denoting a small and red a large size).
\end{example}

Obviously, the previous example is specifically constructed to show the extremes---there is a one-to-one correspondence between gates from $G$ and $G'$ and, hence, an easy way to define an \enquote{optimal} strategy.
In practice, i.e., when $G'\neq G$, no such natural correspondence exists and, as a consequence, it is generally hard to determine an \enquote{optimal} strategy for conducting the simulation.
While this is hardly surprising, given that equivalence checking of quantum circuits has been shown to be computationally hard~\cite{janzingNonidentityCheckQMAcomplete2005}, it underpins the importance of efficient and automated methods for determining suitable simulation paths.

\subsection{Related Work}\label{sec:related_work}

Decision diagrams are not the only data structure for simulating quantum circuits that suffers from the exponential difference in best and worst case complexity.
The connection to tensor networks~\cite{jozsaSimulationQuantumCircuits2006, markovSimulatingQuantumComputation2008, biamonteTensorNetworksNutshell2017} has already been pointed out in Section~\ref{sec:introduction}. In general, a tensor can be understood as a multi-dimensional array of complex numbers---the tensor's rank being the number of dimensions (or indices) while its shape specifies the number of elements in each dimension. 
Two tensors sharing common indices can be \emph{contracted} into a single tensor by summing over repeated indices.
A \emph{tensor network} is a countable set of tensors connected by shared indices. 

\begin{example}\label{ex:matmultensor}
	Let $A, B, C$ be matrices in $\mathbb{C}^{N\times N}$. Further, let the matrix product $C=AB$ be given by
	\begin{equation*}
	{C_{i,j} = \sum_{k=0}^{N-1} A_{i,k}B_{k,j}},
	\end{equation*}
	with $i,j=0,\dots,N-1$.
	Then, this corresponds to the contraction of the rank-$2$ tensors $A = [A_{i,k}]$ and $B = [B_{k,j}]$ over the shared index $k$. 
	This is conveniently represented graphically as:
	\begin{center}
		\begin{tikzpicture}[scale=1.2, every node/.style={scale=1.2}]
			\node[draw] (C) {$C$};
			\node[draw, right = 3 of C] (A) {$A$};			
			\node[draw, right = 1 of A] (B) {$B$};
			\draw[] (C) -- ++ (-0.5, 0) node[left] {$i$};
			\draw[] (C) -- ++ (0.5, 0) node[right] {$j$};
			\draw[] (A) -- ++ (-0.5, 0) node[left] {$i$};
			\draw[] (B) -- ++ (0.5, 0) node[right] {$j$};
			\draw[] (A) -- (B) node[above, midway] {$k$};
			\node[] at ($(C)!0.5!(A)$) {$=$};			
		\end{tikzpicture}
	\end{center}
\end{example}

The order in which a tensor network is contracted into a single tensor is called a \emph{contraction plan}.
In general, an efficient contraction plan tries to keep the size of intermediate tensors and the dimension of contracted indices in check. 
However, the problem of determining an optimal order of contractions has been proven to be NP-hard~\cite{chi-chungOptimizingClassMultidimensional1997}. 
Accordingly, trying to efficiently solve this challenging task for tensor networks has been a heavily researched topic for years~\cite{grayHyperoptimizedTensorNetwork2021, huangClassicalSimulationQuantum2020, boixoSimulationLowdepthQuantum2018, lykovTensorNetworkQuantum2020}.

Both techniques---decision diagrams and tensor networks---efficiently represent the initial state as well as all the individual operations in the form of a dedicated data structure.
Then, they choose a certain path to combine these individual descriptions in order to eventually form a representation of the final quantum state---either by multiplying decision diagrams or by contracting tensors.
Hence, the problem of determining an optimal simulation path for decision diagrams poses a similar challenge as determining an optimal contraction order for a tensor network.

However, in case of decision diagrams, this question is hardly studied and almost no heuristics exist for determining an efficient simulation path.
Initial works related to the problem considered in this work have been conducted in~\cite{zulehnerMatrixVectorVsMatrixMatrix2019, burgholzerEfficientConstructionFunctional2021}. 
In~\cite{zulehnerMatrixVectorVsMatrixMatrix2019}, it is shown that initially constructing the functionality of certain building blocks in prominent quantum algorithms such as Grover's~\cite{groverFastQuantumMechanical1996} or Shor's~\cite{shorPolynomialtimeAlgorithmsPrime1997} algorithm (by conducting, potentially expensive, matrix-matrix multiplications) can lead to significant runtime improvements compared to the sequential matrix-vector multiplication approach described above.
In~\cite{burgholzerEfficientConstructionFunctional2021}, the authors describe schemes for constructing the functionality of such building blocks in a more efficient fashion, which can be interpreted as very specific simulation paths. 
But still, only a significantly limited subset of the immense space of possibilities for very specific problems has been explored.

Besides the sizeable difference of research conducted on tensor networks compared to decision diagrams in general, one of the main reasons for this disparity in focusing on the order of execution during simulation can be identified from the fundamental properties of both techniques:
The performance when contracting tensors only depends on their size and shape---not on the actual content (the data) in the tensors.
On the one hand, this implies that an a-priori estimate of a particular contraction plan's performance can be efficiently inferred from the sizes and shapes of the tensors involved in all contractions.
On the other hand, this also means that, without a proper contraction plan, there is nothing to be gained by employing tensor networks.

In contrast, decision diagrams explicitly try to exploit redundancies in the underlying representations rather relying on \enquote{external} characteristics.
As discussed in~\autoref{sec:decision_diagrams}, this allowed them to efficiently represent and simulate even large quantum systems in many cases.
Consequently, while tensor networks are in dire need of efficient contraction plans to achieve peak performance, simulating circuits in a sequential fashion using decision diagrams has been \enquote{good enough} in many cases.

First steps towards combining tensor networks and decision diagrams have been taken in~\cite{hongTensorNetworkBased2020}. 
There, the authors use a variation of decision diagrams as illustrated in \autoref{sec:decision_diagrams} to represent individual tensors in a more efficient fashion, i.e., they show what can be learned from decision diagrams in order to improve tensor networks. 
In contrast, this work investigates whether knowledge from the tensor network domain can lead to improvements for simulation based on decision diagrams.

\section{A Simulation Path Framework}\label{sec:framework}
In an effort to foster the understanding and development of simulation path heuristics for quantum circuit simulation based on decision diagrams, this section presents an \mbox{open-source} simulation path framework.  
This framework allows to execute arbitrary simulation paths via the powerful TaskFlow~\cite{huangTaskflowLightweightParallel2021} library.
Instead of reinventing the wheel and trying to compensate for years of research on tensor contraction methods, the framework includes a push-button flow to employ existing techniques from the tensor network domain while simultaneously providing the means to easily realize dedicated simulation path strategies for decision diagrams. In the following, we describe how the framework itself handles simulation paths and, afterwards, describe the flow for translating the problem from the domain of decision diagrams to the tensor network domain and back again.

\subsection{Handling Simulation Paths}\label{sec:paths}
The simulation of a quantum circuit \mbox{$G = g_1 \dots g_{|G|}$} with the initial state $\ket{\varphi}$ entails the computation of the expression
\begin{equation*}
U_{|G|} * \cdots * U_1 * \ket{\varphi}.
\end{equation*} 
Initially, this requires the construction of decision diagrams for the initial state and the individual gates.
Then, each multiplication in the above expression can be regarded as a \emph{task} that takes two decision diagrams and returns the result of their multiplication.
Thus, a path for the simulation of $G$ corresponds to a sequence of (multiplication) tasks that eventually results in the final state vector.
It is natural to represent such a sequence as a \emph{task dependency graph}.
An example illustrates the idea.

\begin{figure}[t]
\centering
\includegraphics[width=0.68\linewidth]{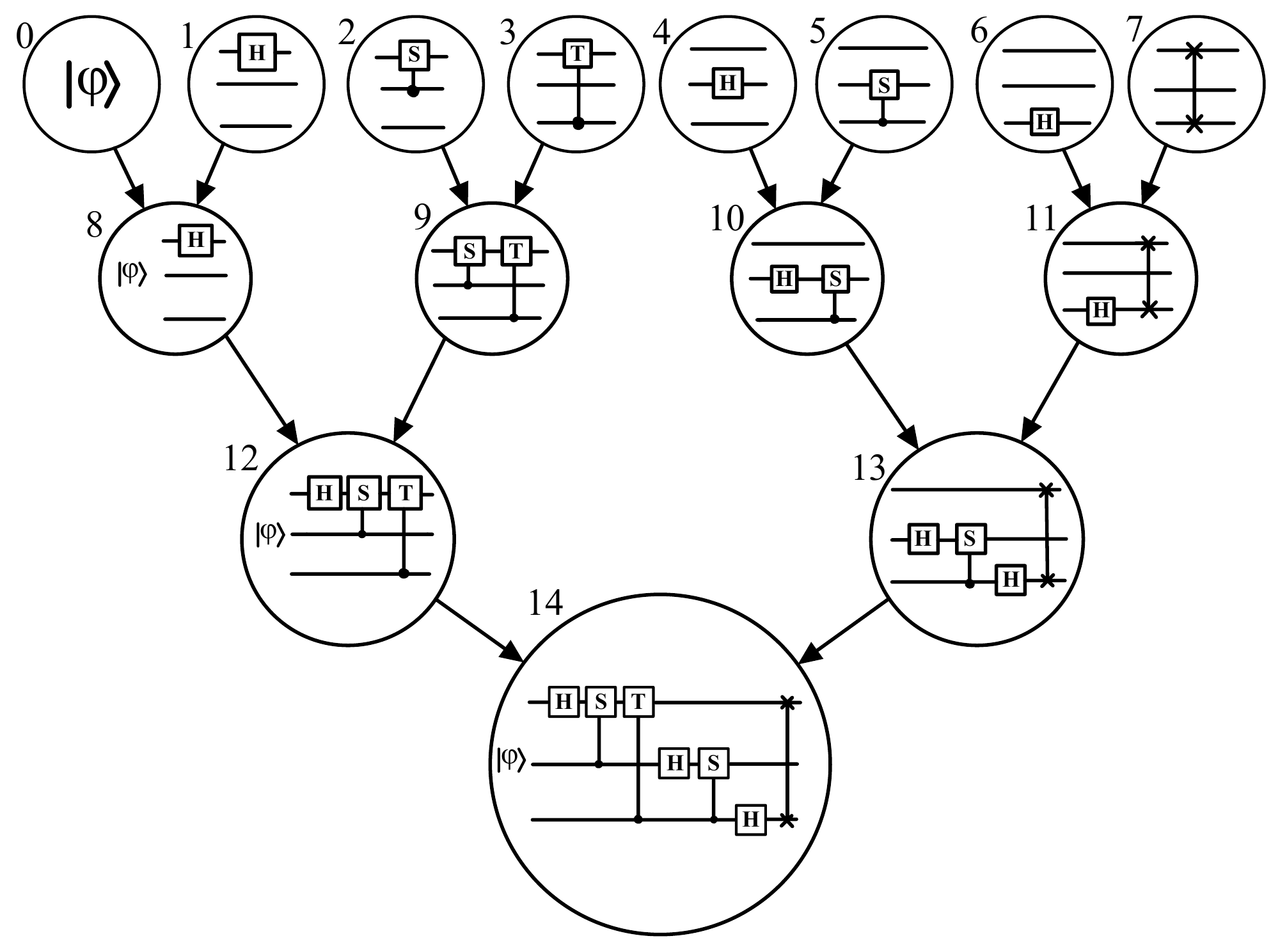} 
\vspace*{-4mm}
\caption{Task dependency graph for one particular simulation path}
\label{fig:task_based_sim_strategie1}
\vspace*{-3mm}
\end{figure}

\begin{figure*}
\centering
\begin{subfigure}[b]{0.66\linewidth}
\centering
\resizebox{0.99\linewidth}{!}{
		\begin{tikzpicture}[style={ultra thick}]
			
			\node[draw, minimum height = 0.8cm] (sv) at (0,-1.5) {$\begin{bNiceMatrix}
\alpha_{000} \\ \alpha_{001} \\ \alpha_{010} \\ \alpha_{011} \\ \alpha_{100} \\	\alpha_{101} \\ \alpha_{110} \\ \alpha_{111}
\end{bNiceMatrix}
$};
			\node (H) at (2,1) [circle, draw] {$\tfrac{1}{\sqrt{2}}\begin{bNiceMatrix}[small]1 & 1\\ 1 & -1\end{bNiceMatrix}$};
			\node (CSA) at (4.5,0.5) [circle, draw]  {$\begin{bNiceMatrix}
				1 &  &  &  \\  & 1 &  &  \\  &  & 1 &  \\  &  &  & i
			\end{bNiceMatrix}$};
			\node (CT) at (8,0.5) [circle, draw]  {$\begin{bNiceMatrix}
				1 &  &  &  \\  & 1 &  &  \\  &  & 1 &  \\  &  &  & \sqrt{i}
			\end{bNiceMatrix}$};
			
			\node[circle, draw, below = 1 of CT] (HA) {$\tfrac{1}{\sqrt{2}}\begin{bNiceMatrix}[small]1 & 1\\ 1 & -1\end{bNiceMatrix}$};	
			
			\node (CSB) at (11.5,-1.5) [circle, draw]  {$\begin{bNiceMatrix}
				1 &  &  &  \\  & 1 &  &  \\  &  & 1 &  \\  &  &  & i
			\end{bNiceMatrix}$};
			
			\node (HB) at (14,-0.5) [circle, draw] {$\tfrac{1}{\sqrt{2}}\begin{bNiceMatrix}[small]1 & 1\\ 1 & -1\end{bNiceMatrix}$};
			
			\node (swap) at (16.5,0.5) [circle, draw]  {$\begin{bNiceMatrix}
				1 &  &  &  \\  &  & 1 &  \\  & 1 &  &  \\  &  &  & 1
			\end{bNiceMatrix}$};
			
			\node[right = 10 of H] (q2){};
			\node[right = 2 of HB] (q0){};
			\node[right = 2 of HA] (q1){};
			\coordinate(svtre) at (0.7,-4);
			\coordinate(svtwo) at (0.7,-1.5);
			\coordinate(svone) at (0.7,1);
			\coordinate(csaone) at (intersection 1 of CSA and q2--H);
			\draw (H) -- (csaone);
			\draw (svtre) -- ++(4.0, 0) to[out=0, in=-170] (CT);
			\draw (svone) to[out=0,in=180] (H);
			\draw (svtwo) to[out=0,in=-170] (CSA);
			\coordinate(csatwo) at (intersection 1 of CSA and H--q2);
			\coordinate(ctone) at (intersection 1 of CT and q2--H);
			\draw (csatwo) -- (ctone);
			
			\coordinate(cttwo) at (intersection 1 of CT and H--q2);
			\coordinate(swapone) at (intersection 1 of swap and q2--H);
			\draw (cttwo) -- (swapone);
			
			\coordinate(csbone) at (intersection 1 of CSB and q1--HA);
			\draw (HA) to[in=-170, out=0] (CSB);

			\draw (CSA) to[out=-10,in=180] (HA);
			\draw (CT) to[out=-10,in=170] (CSB);
			\draw (HB) to[out=0,in=180] (swap);
			\draw (CSB) to[out=10,in=180] (HB);
			\draw (CSB.-10) -- ++ (6, 0);
			
			\coordinate(swaptwo) at (intersection 1 of swap and H--q2);
			\coordinate(swaptre) at (19,-4);
			\draw[] (swaptwo) -- ++ (1, 0);
			\draw (swap) to[out=0, in=180] (swaptre); 	
		\end{tikzpicture}}
	\caption{Tensor network}
	\label{fig:tn_in_flow}
	\end{subfigure}%
	\hfill
	\begin{subfigure}[b]{0.33\linewidth}
	\centering
	\includegraphics[width=0.99\linewidth]{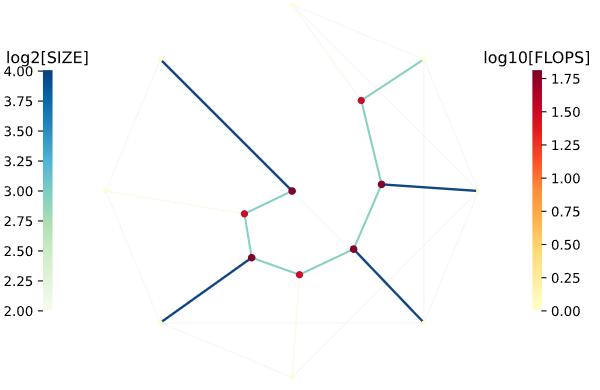}
	\caption{Contraction plan visualization}
	\label{fig:contractionring}
\end{subfigure}
\caption{Tensor network for the simulation of the 3-qubit QFT and visualization of a corresponding contraction plan }
\label{fig:tncircandvis}
\end{figure*} 

\begin{example}\label{ex:simpath}
	Consider again the $3$-qubit QFT circuit shown in \autoref{fig:circuit}.
	Then, the following sequence of tasks describes one particular simulation path of $G$: 
	\[
	[(0,1), (2, 3), (4, 5), (6, 7), (8, 9), (10, 11), (12, 13)].
	\]
	To this end, index $0$ denotes the initial state, index $1$ to $|G|$ the individual operations, and the result of a task is indexed by the next largest integer not already in use.
	The corresponding graph is shown in \autoref{fig:task_based_sim_strategie1}.
\end{example}

This task-based formulation of quantum circuit simulation allows to employ powerful tools for asynchronous task-parallelism~\cite{huangTaskflowLightweightParallel2021, carteredwardsKokkosEnablingManycore2014, kaiserHPXStandardLibrary2020} to conduct the simulation.
The resulting framework takes a circuit and a sequence of tasks as an input and uses the Taskflow~\cite{huangTaskflowLightweightParallel2021} library to build the corresponding task dependency graph and execute it asynchronously\footnote{A similar effort has been conducted for the tensor network domain in~\cite{vincentJetFastQuantum2021}.}.
The question remains how to determine suitable ones out of the $|G|!$ options.

\subsection{Utilizing Research on Tensor Network Contraction}\label{sec:flow}

As reviewed in \autoref{sec:related_work}, a plethora of methods has been developed for determining efficient contraction plans for tensor networks and first steps have been taken to combine these two techniques. 
Due to the direct connection between the two domains, it would not make sense to try and reinvent the wheel when it comes to simulation using decision diagrams.
Instead, we realized a flow in Python that connects both domains and, as a consequence, allows to make use of research conducted towards tensor network contraction. 
The following (rather technical) paragraphs give a detailed description of this process.

Starting from an initial quantum circuit (provided in the form of an \emph{OpenQASM} file~\cite{crossOpenQASMBroaderDeeper2021} or Qiskit \mbox{\emph{QuantumCircuit}} object~\cite{aleksandrowiczQiskitOpensourceFramework2019}), the first step is to create a corresponding tensor network representation.
To this end, each individual gate is transformed to a corresponding tensor representing the underlying matrix. 
We do not employ tensor slicing techniques (as demonstrated in~\cite{chen64qubitQuantumCircuit2018, pednaultParetoefficientQuantumCircuit2020, huangClassicalSimulationQuantum2020}), which allow to split the tensors of multi-qubit gates into multiple smaller tensors, since these techniques are not yet widely adopted in the decision diagram domain (although first efforts towards this direction have been conducted in~\cite{burgholzerHybridSchrodingerFeynmanSimulation2021}).

Next, the initial state $\ket{\varphi}$ needs to be translated to the tensor network domain.
In general, an $n$-qubit state is described by a rank-$n$ tensor of size $2^n$, i.e., the complete state vector. 
In case of product states, i.e., states that can be written as a product of single-qubit states such as the all-zero state $\ket{0\dots 0} = \ket{0}\otimes\dots\otimes\ket{0}$, this can rather be represented as $n$ rank-$1$ tensors of size $2$.
A similarly compact representation is achieved for decision diagrams of product states, which always consist of $n$ nodes (as opposed to the general worst case of $2^{n-1}$ nodes).
However, while tensor networks allow for arbitrary contractions between two tensors as long as they share a common index, decision diagrams for quantum computing, as considered in this work, do not support arbitrary kinds of (tensor) contractions as of now.
Instead, they only support (proper) matrix-vector and matrix-matrix multiplication, i.e., it is, for example, not possible to contract the (vector) decision diagram representing a single-qubit state and the (matrix) decision diagram representing a two-qubit operation.
As a consequence, the initial state in the translated tensor network needs to be represented as a full rank-$n$ tensor (see \autoref{sec:discussion} for further discussions).

Contracting the resulting tensor network results in a tensor representing the complete output state vector of the simulation.

\begin{example}\label{ex:tnexample}
	Consider once more the $3$-qubit QFT circuit shown in \autoref{fig:circuit}. 
	Then, the corresponding tensor network representation for the simulation of the circuit is shown in~\autoref{fig:tn_in_flow}.
\end{example}

After the translation, the resulting tensor network can be fed into any available tensor network contraction tool in order to determine a suitable contraction plan. 
We used the \mbox{hyper-optimized} tensor network contraction tool \mbox{CoTenGra}~\cite{grayHyperoptimizedTensorNetwork2021} as a state-of-the-art representative. 
It allows to determine contraction plans for large tensor networks using various graph-based methods and is publicly available at \url{github.com/jcmgray/cotengra}.
Furthermore, it provides means to visualize contraction plans and their complexity in meaningful ways.

\begin{example}\label{ex:cotengra}
	Feeding the tensor network shown in \autoref{fig:tn_in_flow} into CoTenGra results in the following simulation path:
	\[
	[(0, 1), (2, 8), (3, 9), (4, 10), (5, 11), (6, 12), (7, 13)].
	\]
	The resulting contraction tree can be visualized as shown in \autoref{fig:contractionring}.
	To this end, the color of each node represents the number of floating point operations required for a particular contraction, while the color of each edge represents the size of the respective tensors---the darker the color, the more complex the contraction or the larger the tensor.
\end{example}

\begin{figure}[t]
\centering
\includegraphics[scale=0.155]{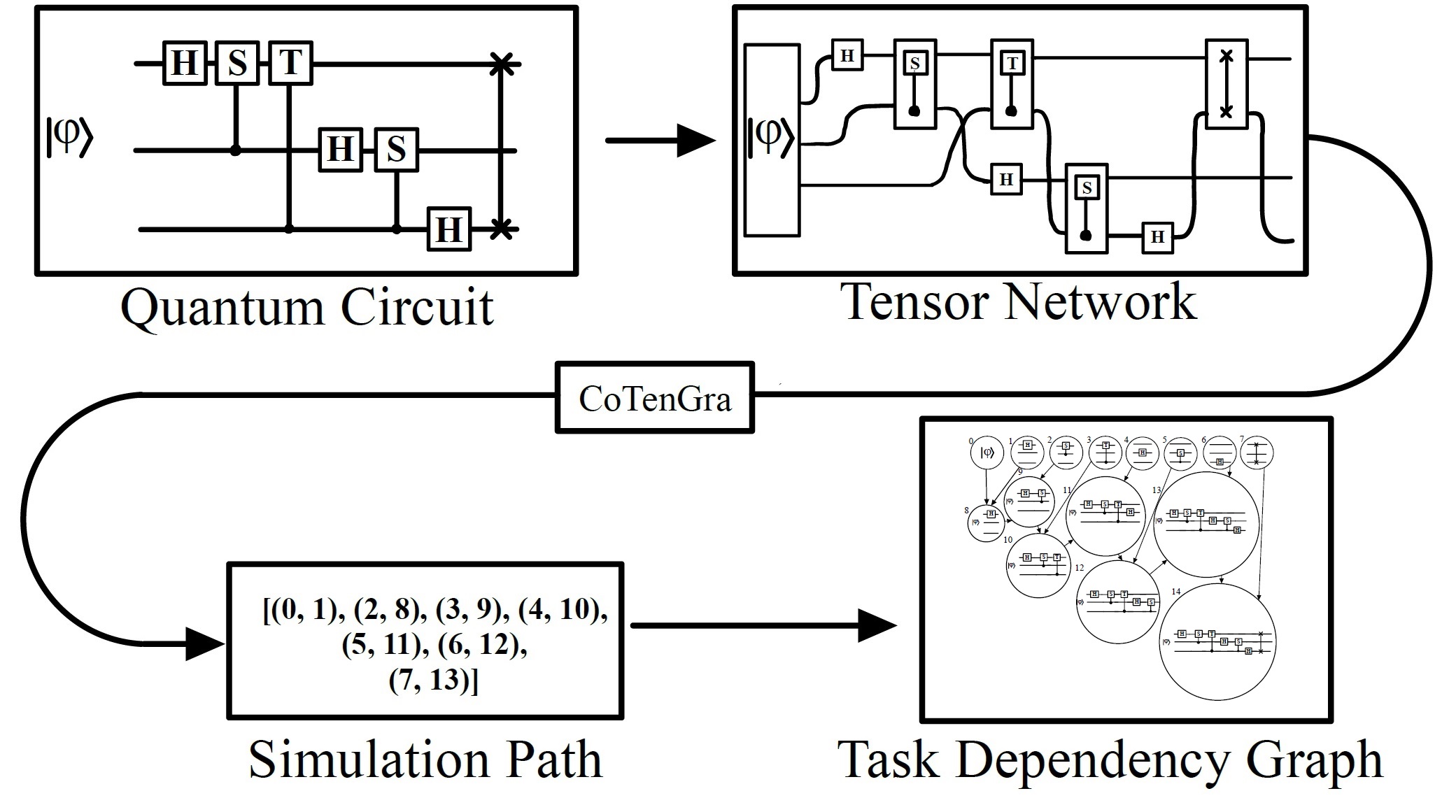}
\caption{Automated simulation path flow inspired by the tensor network domain}
\label{fig:flow_example}
\end{figure}

Overall, this results in a flow as shown in \autoref{fig:flow_example}, where the initial quantum circuit is first translated to a tensor network and then fed into CoTenGra. Afterwards, a task dependency graph is constructed from the obtained contraction plan and used for the decision diagram simulation.

\section{Experimental Evaluations}
\label{sec:results}

The proposed simulation path framework has been implemented on top of the publicly available decision diagram-based simulator \emph{DDSIM} (available at \url{github.com/cda-tum/ddsim}), which is part of the \emph{Munich Quantum Toolkit} (MQT, formerly known as JKQ~\cite{willeJKQJKUTools2020}).
Afterwards, we used the resulting tool to conduct an extensive case study in order to $(1)$ evaluate the effectiveness of strategies developed in the tensor network domain, and $(2)$ determine the potential of dedicated heuristics for quantum circuit simulation using decision diagrams.
All evaluations have been conducted on a machine equipped with an \mbox{AMD Ryzen 9 3950X} CPU and \SI{128}{\gibi\byte} RAM running Ubuntu 20.04. \pagebreak

\subsection{Experimental Setup}\label{sec:setup}

Inspired by the discussions in \autoref{sec:considered_problem}, we primarily focused on the typical use case of verifying the equivalence of two quantum circuits, i.e., for two given quantum circuits $G$ and $G'$, we considered the simulation of the combined circuit $\tilde{G} = G G^{\prime -1}$.
In order to create meaningful verification instances, a broad selection of circuits $G$ and $G'$ has been taken from the publicly available benchmark suite MQT Bench~\cite{quetschlichMQTBenchBenchmarking2022}, which offers various quantum algorithms on different abstraction levels. 
More specifically, the circuits $G$ have been taken from the \emph{algorithmic layer} (the highest available abstraction), while the circuits $G'$ have been taken from the \emph{native-gates layer} (where circuits are compiled and optimized for a particular architecture). As discussed in~\autoref{sec:considered_problem}, this creates a non-trivial verification scenario where there no longer is a one-to-one correspondence between the gates of $G$ and $G'$ (as, e.g., shown in~\autoref{ex:worstcasesimpath}).

\subsection{Dedicated Simulation Path Heuristic}\label{sec:heuristic}

In order to evaluate the potential of a non-trivial, dedicated simulation path scheme, we developed and implemented a heuristic that aims to efficiently solve this kind of verification tasks by exploiting some knowledge about the compilation flow itself (inspired by the ideas in~\cite{burgholzerVerifyingResultsIBM2020}).
For each gate~$g$ in the original circuit $G$, the resulting method estimates the corresponding number of operations in $G^{\prime}$ based on the decomposition of $g$ into the native gate-set of $G'$ (which can easily be computed a-priori using the same settings for the decomposition as were used for compiling $G$ to $G'$).
Then, starting in between both circuits, any application of a gate from $G$ is followed by the application of the corresponding number of gates from $G^{\prime -1}$, until only the final multiplication with the initial state vector remains---yielding the final result of the simulation.

In all but the most trivial cases, any method based on this principle can only ever derive an approximation of the actual number of operations in the compiled circuit, since already the simplest optimizations employed during compilation frequently eliminate a significant amount of gates from the overall circuit.
However, as witnessed by the evaluations, this type of simulation path still frequently allows to keep the intermediate decision diagrams throughout the simulation very compact (i.e., close to the identity structure) and, hence, allows for an efficient simulation.

\subsection{Experimental Results}

A representative subset of the obtained results is summarized in \autoref{tab:results}---with the first three columns denoting the name of the benchmark, the number of qubits $n$, as well as the number of gates $|\tilde{G}|$ of the combined circuit. 
The remaining columns show
\begin{itemize}
	\item the runtime $t_\mathit{seq}$ of the state-of-the-art, i.e., sequential, simulation path strategy~\cite{zulehnerAdvancedSimulationQuantum2019},
	\item the runtime $t_\mathit{ten}$ of the proposed flow for translating strategies from the tensor network domain (split into the time $t_\mathit{cot}$ spent on determining a simulation path using CoTenGra\footnote{The tool allows to set a timeout and a maximum number of repetitions, which were set to \SI{600}{\second} and $65$, respectively.} and the simulation time $t_\mathit{sim}$), as well as
	\item the runtime $t_\mathit{heur}$ of the proposed dedicated simulation path heuristic.
\end{itemize}
The framework as well as the benchmark script are publicly available at \url{github.com/cda-tum/ddsim} to conduct further evaluations.

In a first series of evaluations, we used the design flow proposed in \autoref{sec:flow} to make use of the plethora of available tensor network strategies to determine a suitable simulation path.
As shown by these results, re-using or translating methods developed in the tensor network domain via the proposed flow can already speed up the simulation of quantum circuits using decision diagrams by a large margin compared to the state-of-the-art, i.e., sequential, approach~\cite{zulehnerAdvancedSimulationQuantum2019} (runtimes of instances where $t_{\mathit{ten}} < t_{\mathit{seq}}$ are highlighted in \textbf{bold}).
Interestingly, for some cases (such as the \emph{Graph State} benchmark on $54$ qubits), the substantially improved runtime ($t_\mathit{sim}$) is overshadowed by the time spent on searching for a suitable path ($t_\mathit{cot}$)---these instances are highlighted in \textit{italic}.  
In other cases, nothing at all is to be gained by spending up to \SI{600}{\second} on the search for a suitable path. 
This is further discussed in~\autoref{sec:discussion}. 

In a second series of evaluations, we studied the performance of the dedicated heuristic simulation path method proposed above.
The results clearly underline the potential of dedicated simulation path schemes for decision diagrams.
Although the proposed method is only a heuristic, it oftentimes yields several orders of magnitude faster runtimes compared to the state of the art (again, runtimes where $t_\mathit{heur} < t_\mathit{seq}$ are highlighted in \textbf{bold}).
Most notably, for the \emph{Entangled QFT} benchmark on $18$ qubits, choosing the right simulation path makes the difference between waiting almost two hours for the result and having it available in the blink of an eye.

\begin{table}[t]
	\centering
	\caption{Experimental Evaluations}\label{tab:results}
	\includegraphics[width=0.95\linewidth]{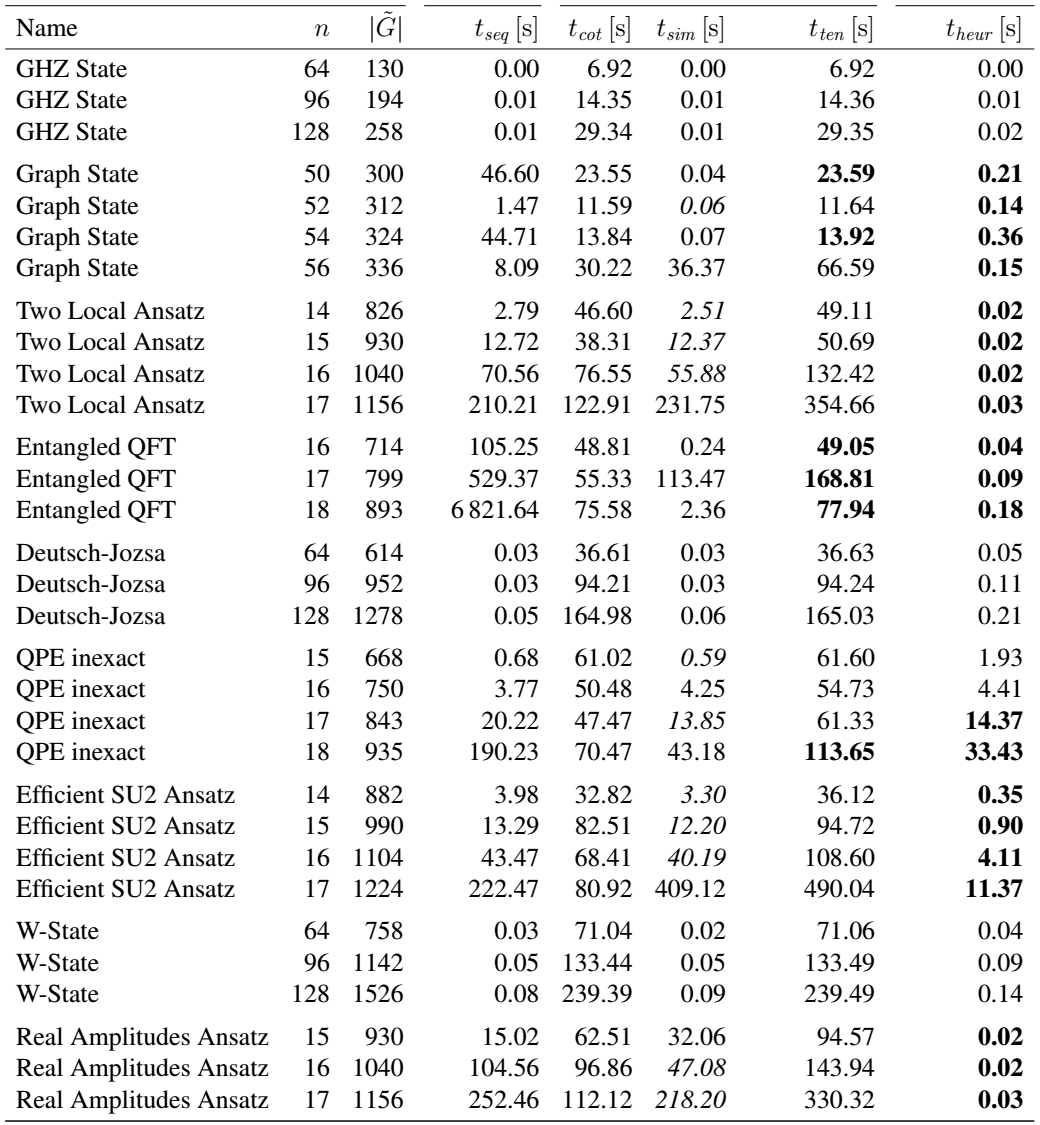}
	\\\vspace{2mm}
	{\small $n$: Number of qubits \hspace*{0.4cm} \emph{$\vert G \vert$}: Gate count of $G$ \\ $t_{\mathit{seq}}$: Runtime of sequential simulation path~\cite{zulehnerAdvancedSimulationQuantum2019} \\ $t_{\mathit{cot}}$: Runtime of path determined by CoTenGra~\cite{grayHyperoptimizedTensorNetwork2021} \\ $t_{\mathit{opt}}$: Runtime of optimal path }
\vspace*{-0.5cm}
\end{table}

\section{Discussion}
\label{sec:discussion}

While the problem of finding a suitable simulation path for decision diagrams has many apparent similarities to finding an efficient contraction plan for tensor networks, the results summarized above demonstrate that there are key differences between both data structures.
Based on that, interesting conclusions can be drawn on what \emph{can} be learned from tensor networks and what \emph{cannot} for quantum circuit simulation based on decision diagrams.

For tensor networks, it is inevitable that the final result of the full state simulation is exponentially large. 
This inherently limits the usability of such simulations to the available system memory.
In contrast, there are many known quantum algorithms whose state vectors during the simulation emit a very compact (in the best case, linear) decision diagram representation~\cite{zulehnerAdvancedSimulationQuantum2019} (also witnessed in the evaluation above for the \emph{GHZ State}, \emph{W-State}, and \emph{Deutsch-Jozsa} benchmarks).
As such, there is no strict qubit limit for simulation based on decision diagrams.
Furthermore, the margin between the simulation times of all three considered strategies is negligible and, hence, it does not make sense to spend any amount of time searching for an alternative simulation path in these cases.

One might be tempted to conclude that, given the same simulation path/contraction plan, simulation based on decision diagrams always has at most as large a memory footprint and takes at most as long as the corresponding tensor network contraction---which would imply that decision diagrams are strictly superior to tensor networks.
However, in order to maintain the decision diagram structure, a data structure more advanced than simple multi-dimensional arrays (as commonly used for tensors) is necessary.
As a consequence, each node in a decision diagram is significantly larger in memory than individual tensor entries. 
Since, in the worst case, the decision diagram consists of exponentially many unique nodes ($2^{n-1}$ for an $n$-qubit state), the simulation might incur a much higher memory footprint.
In addition, decision diagram operations frequently make use of compute tables in order to avoid redundant computations and exploit any redundancy present in the representations.
If the decision diagram contains no redundancies, all that constitutes a significant overhead compared to straight-forward tensor contraction (as, e.g., observed for the sequential simulation of the \emph{Entangled QFT} benchmark).
Overall, decision diagrams and tensor networks complement each other in many different ways and, therefore, should be employed under consideration of the specific task at hand.

Although finding a suitable simulation path inspired by the tensor network domain may consume a significant amount of time (in some cases overshadowing the subsequent runtime improvements), the runtime ($t_\mathit{sim}$) of the simulation paths determined by CoTenGra demonstrates that there \emph{is} something to be learned from the domain of tensor networks. 
However, the current state of the art in decision diagrams for quantum computing, as they are considered in this work, imposes several limitations on what can actually be learned from the domain of tensor networks.

One of the main (technical) restrictions of decision diagrams \emph{at the moment} is that they only allow for proper matrix-vector and matrix-matrix multiplication while tensor networks allow for arbitrary contractions between two tensors (as long as they share a common index).
So, while the following contraction between a single-qubit state and a two-qubit operation makes perfect sense for tensor networks
\begin{center}
	\begin{tikzpicture}[scale=0.8, every node/.style={scale=0.8}]
		\node[draw, circle, inner sep=1pt, fill=gray!20] (q) {$\ket{0}$};
		\node[draw, circle, inner sep=1pt, below=0.1 of q] (p) {$\ket{0}$};
		\node[draw, fit={(p) (q)}, xshift=0.9cm, inner sep=0pt,label=center:U, fill=gray!20] (U) {};
		\draw (p) -- (p -| U.west);
		\draw (q) -- (q -| U.west);
		\draw (p -| U.east) -- ++ (0.5, 0);
		\draw (q -| U.east) -- ++ (0.5, 0);
		
		\node[circle, inner sep=1pt, right=2cm of q] (qq) {\phantom{$\ket{0}$}};
		\node[draw, circle, inner sep=1pt, below=0.1 of qq] (pp) {$\ket{0}$};
		\node[draw, fit={(pp) (qq)}, xshift=0.9cm, inner sep=0pt,label=center:$U'$, fill=gray!20] (UU) {};
		\draw (pp) -- (pp -| UU.west);
		\draw (pp -| UU.east) -- ++ (0.5, 0);
		\draw (qq -| UU.east) -- ++ (0.5, 0);
		
		\node[] at ($(U)!0.4!(UU)$) {$=$};			
	\end{tikzpicture},
\end{center}
decision diagrams currently only permit the following type of contraction
\begin{center}
	\begin{tikzpicture}[scale=0.8, every node/.style={scale=0.8}]
		\node[draw, inner sep=1pt, fill=gray!20, align=left] (q) {$\ket{0}$\\$\otimes$\\$\ket{0}$};
		\node[draw, fit={(q)}, xshift=0.9cm, inner sep=1pt,label=center:U, fill=gray!20] (U) {};
		\draw[thick, double] (q) -- (q -| U.west);
		\draw[thick, double] (q -| U.east) -- ++ (0.5, 0);
				
		\node[draw, fit={(q) (U)}, xshift=2.5cm, inner sep=1pt,label=center:$U\ket{00}$, fill=gray!20] (UU) {};
		\draw[thick, double] (UU.east) -- ++ (0.5, 0);

		\node[] at ($(U)!0.5!(UU)$) {$=$};			
	\end{tikzpicture}.
\end{center}
This limits the degrees of freedom to explore during the search for an efficient contraction plan and eliminates some of the benefits of tensor networks as it essentially fixes the tensor of the initial state to be maximally large.

In a similar fashion, so-called tensor slicing techniques have been shown to accelerate and trivially parallelize tensor network contractions~\cite{chen64qubitQuantumCircuit2018, pednaultParetoefficientQuantumCircuit2020, huangClassicalSimulationQuantum2020}.
There, the main idea is to split multi-qubit tensors into multiple smaller tensors of higher order, as exemplary illustrated in the following
\begin{center}
	\begin{tikzpicture}[scale=0.8, every node/.style={scale=0.7}]
		\node[draw, circle, inner sep=1pt] (q) {$\ket{0}$};
		\node[draw, circle, inner sep=1pt, below=0.1 of q] (p) {$\ket{0}$};
		\node[draw, fit={(p) (q)}, xshift=0.9cm, inner sep=0pt,label=center:U, fill=gray!20] (U) {};
		\draw (p) -- (p -| U.west);
		\draw (q) -- (q -| U.west);
		\draw (p -| U.east) -- ++ (0.5, 0);
		\draw (q -| U.east) -- ++ (0.5, 0);
		
		\node[draw, circle, inner sep=1pt, right=2cm of q] (qq) {$\ket{0}$};
		\node[draw, circle, inner sep=1pt, below=0.1 of qq] (pp) {$\ket{0}$};
		\node[draw, fit={(q)}, inner sep=0pt, fill=gray!20, right=0.4cm of qq, label=center:$U_1$] (Uq) {};
		\node[draw, fit={(p)}, inner sep=0pt, fill=gray!20, right=0.4cm of pp, label=center:$U_0$] (Up) {};
		\draw (pp) -- (pp -| Up.west);
		\draw (qq) -- (qq -| Uq.west);
		\draw (pp -| Up.east) -- ++ (0.5, 0);
		\draw (qq -| Uq.east) -- ++ (0.5, 0);
		\draw (Uq.south) -- (Up.north);
		
		\node[] at ($(U)!0.4!($(Up)!0.5!(Uq)$)$) {$=$};			
	\end{tikzpicture}.
\end{center}
While first efforts towards employing such techniques for decision diagrams have been conducted in ~\cite{burgholzerHybridSchrodingerFeynmanSimulation2021}, they are not yet mature and flexible enough to be employed in the same way as for tensor networks.

As discussed above, employing tensor networks for full quantum state simulation inherently carries an exponential memory complexity.
For that reason, tensor networks are commonly rather used to determine individual amplitudes of the resulting state.
This is accomplished by attaching additional tensors to the end of the tensor network that describe the desired amplitude, as exemplary illustrated by
\begin{center}
	\begin{tikzpicture}[scale=0.8, every node/.style={scale=0.7}]
		\node[draw, circle, inner sep=1pt] (q) {$\ket{0}$};
		\node[draw, circle, inner sep=1pt, below=0.1 of q] (p) {$\ket{0}$};
		\node[draw, fit={(p) (q)}, xshift=0.9cm, inner sep=0pt,label=center:U] (U) {};
		\draw (p) -- (p -| U.west);
		\draw (q) -- (q -| U.west);
		
		\node[draw, circle, inner sep=1pt, fill=gray!20, right = 1.1cm of q] (qq) {$\ket{0}$};
		\node[draw, circle, inner sep=1pt, fill=gray!20, right = 1.1cm of p] (pp) {$\ket{0}$};
		
		\draw (p -| U.east) -- (pp);
		\draw (q -| U.east) -- (qq);
		
		\node[right=1cm of U] (eq) {$\stackrel{contract}{\rightarrow}$};
		\node[right=0.2cm of eq, draw, circle, inner sep=1pt] {$\alpha_{00}$};
			
	\end{tikzpicture}.
\end{center}
Since the result of the simulation no longer is an exponentially large vector, but rather a single scalar, the order of contractions plays a far greater role for the efficiency of such simulation.
At the moment, the only established way to determine individual amplitudes using decision diagrams is to compute the decision diagram for the full state vector and, then, extract the desired amplitude by traversing the decision diagram from top to bottom.
While there still are many instances where this procedure works just fine, this imposes a significant restriction on the available degrees of freedom for determining a suitable simulation path.

\section{Conclusions}
\label{sec:conclusions}

In this work, we studied the importance of the path that is chosen when simulating quantum circuits using decision diagrams.
The resulting framework allows to employ arbitrary simulation paths and to connect the domain of tensor networks with the domain of decision diagrams. 
Our experimental evaluations have shown that much can be learned from the domain of tensor networks---potentially allowing for runtime improvements of up to several factors compared to the state of the art.
In addition, we demonstrated that the development of application-specific heuristics which are tailored for decision diagrams can achieve speedups of several orders of magnitude compared to the state of the art.
Finally, we have shown, conceptually as well as experimentally, that decision diagrams and tensor networks differ in some key aspects and that not everything can be learned from tensor networks---at least not given the current state of the art in decision diagrams for quantum computing.
As decision diagrams, which still are a very young data structure compared to tensor networks, are developed further, the potential for taking advantage of all the research conducted towards tensor networks is expected to increase drastically.
By making the developed framework publicly available, we hope to accelerate such endeavors.

\section*{Acknowledgments}
This work received funding from the European Research Council (ERC) under the European Union’s Horizon 2020 research and innovation program (grant agreement No. 101001318), was part of the Munich Quantum Valley, which is supported by the Bavarian state government with funds from the Hightech Agenda Bayern Plus, and has been supported by the BMWK on the basis of a decision by the German Bundestag through project QuaST, as well as by the BMK, BMDW, and the State of Upper Austria in the frame of the COMET program (managed by the FFG).

\printbibliography

\begin{IEEEbiography}
	[{\includegraphics[width=1in,height=1.25in,clip,keepaspectratio]{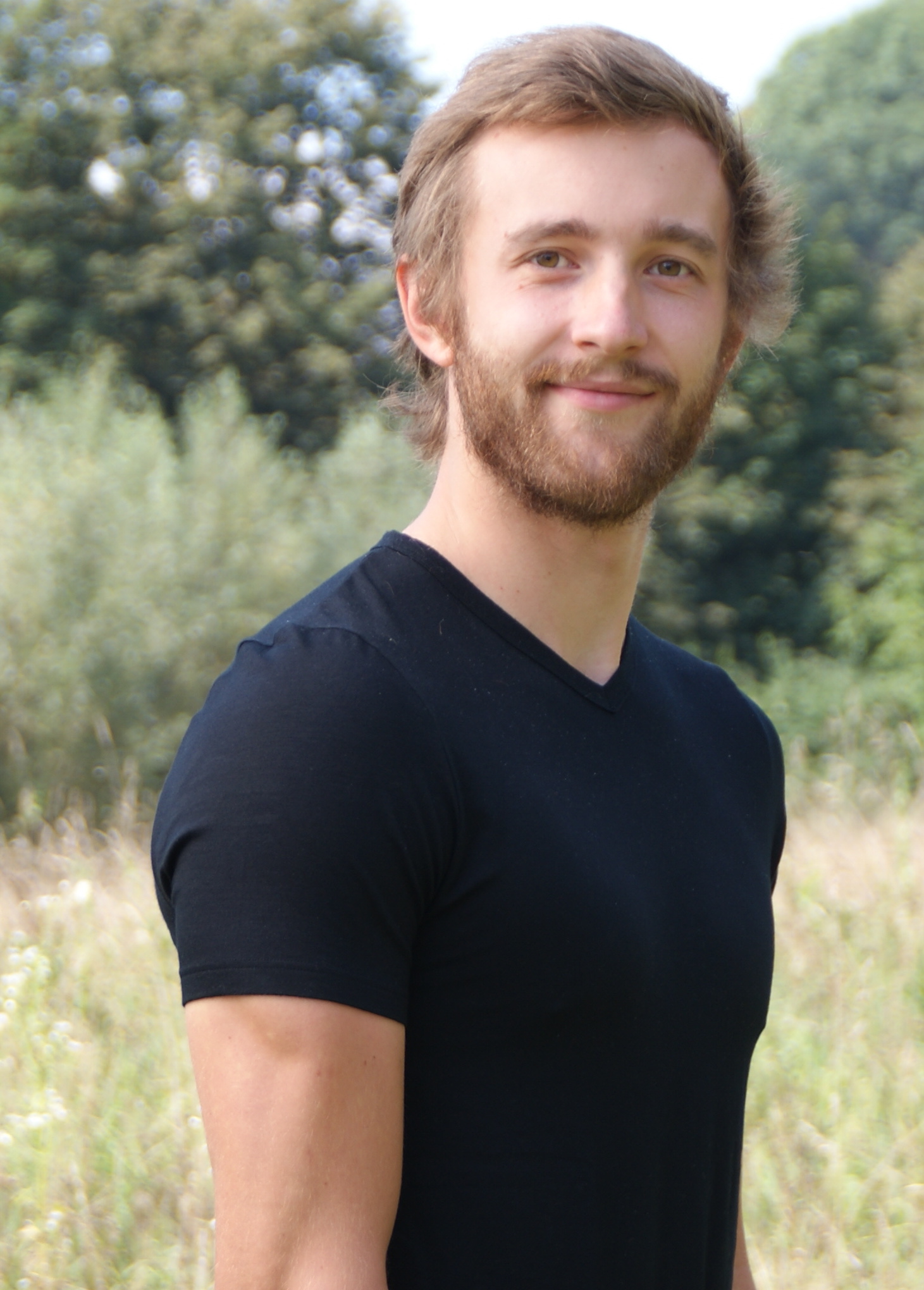}}]{Lukas Burgholzer}
	Lukas Burgholzer (S’19) received his Master's degree in industrial mathematics (2018) and Bachelor's degree in computer science (2019) from the Johannes Kepler University Linz, Austria.
	He is currently a Ph.D. student at the Institute for Integrated Circuits at the Johannes Kepler University Linz, Austria. 
	His research focuses on design automation and software for quantum computing. In these areas, he has published several papers on international conferences such as ASP-DAC, DAC, ICCAD, DATE, and QCE.
\end{IEEEbiography}

\begin{IEEEbiography}
	[{\includegraphics[width=1in,height=1.25in,clip, keepaspectratio]{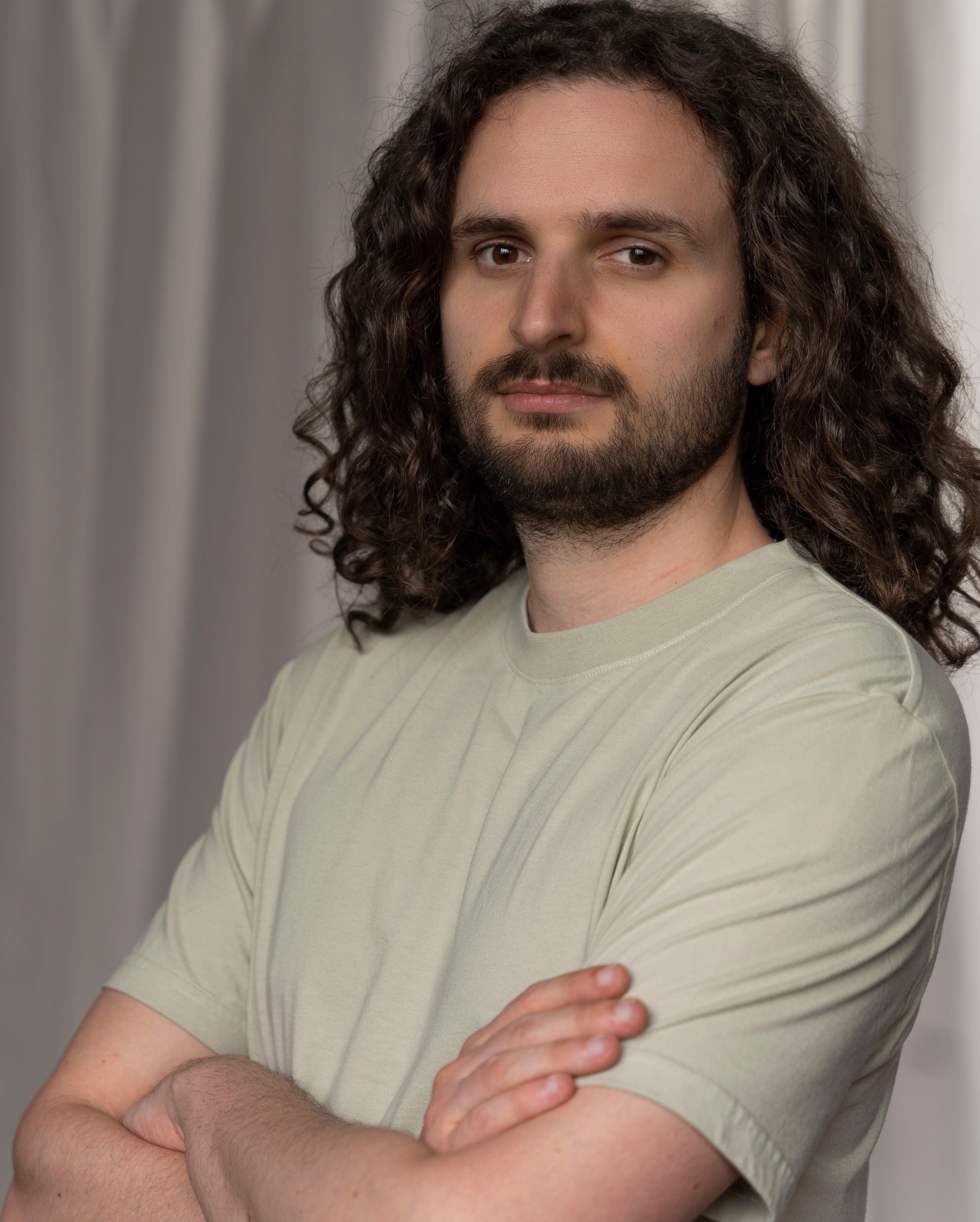}}]{Alexander Ploier}
	Alexander Ploier received his Master's degree in industrial mathematics (2019) from the Johannes Kepler University Linz, Austria.
	He is currently a Ph.D. student at the Institute for Integrated Circuits at the Johannes Kepler University Linz, Austria. 
	His research interests include design automation for quantum computing---currently focusing on different data structures. 
\end{IEEEbiography}

\begin{IEEEbiography}
	[{\includegraphics[width=1in,height=1.25in,clip,keepaspectratio]{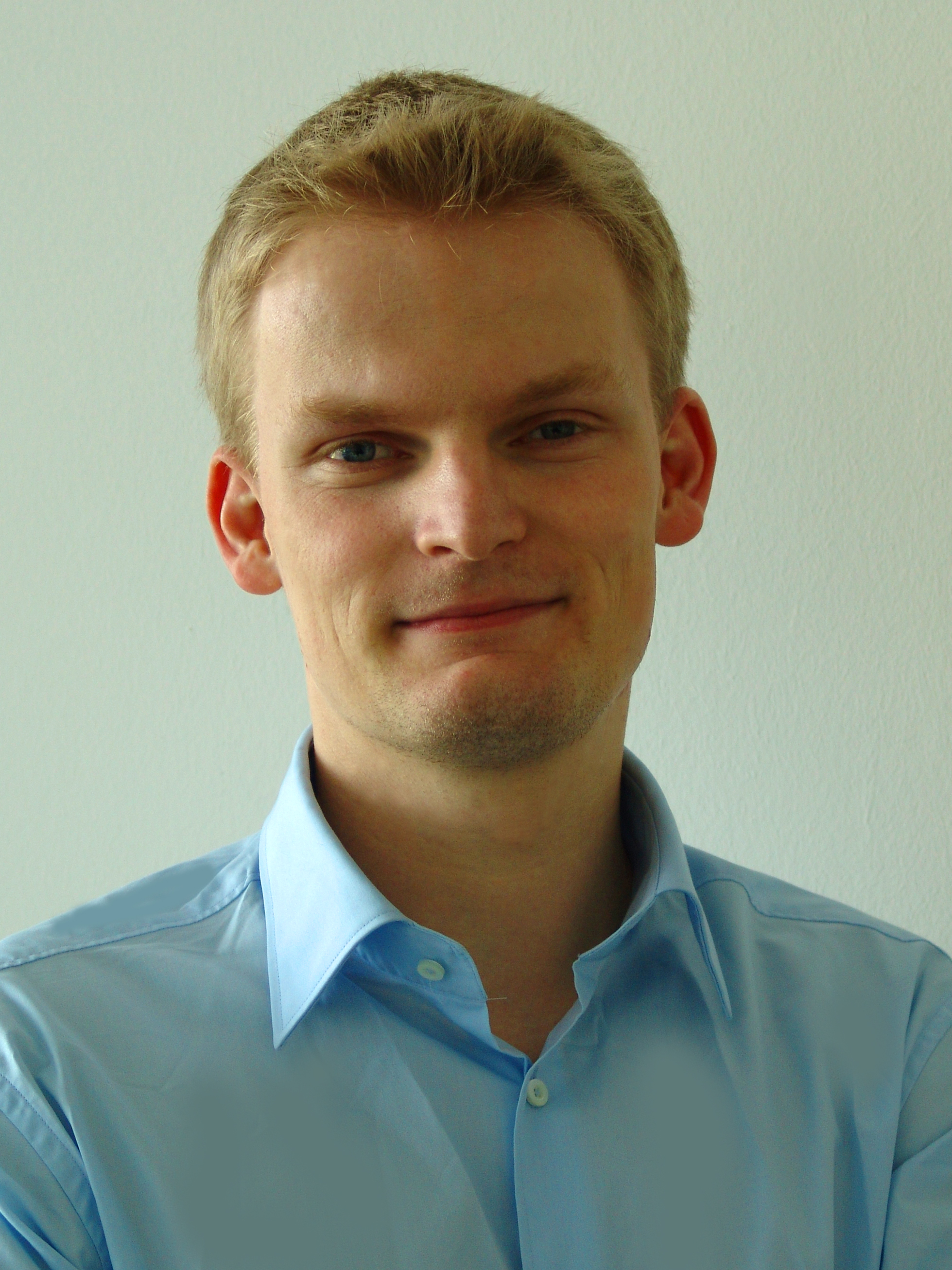}}]{Robert Wille}
Robert Wille is a Full and Distinguished Professor at the Technical University of Munich, Germany, and Chief Scientific Officer at the Software Competence Center Hagenberg, Austria. He received the Diploma and Dr.-Ing. degrees in Computer Science from the University of Bremen, Germany, in 2006 and 2009, respectively. Since then, he worked at the University of Bremen, the German Research Center for Artificial Intelligence (DFKI), the University of Applied Science of Bremen, the University of Potsdam, and the Technical University Dresden. From 2015 until 2022, he was Full Professor at the Johannes Kepler University Linz, Austria, until he moved to Munich. His research interests are in the design of circuits and systems for both conventional and emerging technologies. In these areas, he published more than 400 papers and served in editorial boards as well as program committees of numerous journals/conferences such as TCAD, ASP-DAC, DAC, DATE, and ICCAD. For his research, he was awarded, e.g., with Best Paper Awards, e.g., at TCAD and ICCAD, an ERC Consolidator Grant, a Distinguished and a Lighthouse Professor appointment, a Google Research Award, and more.
\end{IEEEbiography}

\end{document}